\documentclass{eptcs}
\usepackage[utf8]{inputenc}
\usepackage{bussproofs}
\usepackage{graphicx}
\DeclareUnicodeCharacter{22A5}{$\bot$}
\DeclareUnicodeCharacter{22A2}{$\vdash$}
\DeclareUnicodeCharacter{03B5}{$\epsilon$}
\DeclareUnicodeCharacter{21D2}{$\Rightarrow$}
\DeclareUnicodeCharacter{00AC}{$\neg$}
\DeclareUnicodeCharacter{2228}{$\vee$}
\newcommand{\horizontalsize}{.16}
\newcommand{\verticalsize}{.13}

\def\titlerunning{A Mobile Application for Self-Guided Study of Formal Reasoning}
\def\authorrunning{D.M. Cerna, R.P.D Kiesel, \& A. Dzhiganskaya}

\title{A Mobile Application for Self-Guided Study of Formal Reasoning}

\author{David M. Cerna
\institute{Formal Methods and Verification,\ johannes Kepler University \\ Research Institute for Symbolic Computation,\ johannes Kepler University}
\email{David.Cerna@risc.jku.at}
\and
Rafael P.D. Kiesel
\institute{Knowlege Based Systems,\ Technical University of Vienna}
\email{rafael.kiesel@tuwien.ac.at}
\and 
Alexandra Dzhiganskaya
\institute{University of Applied Arts Vienna}
\email{dzhyganska.alexandra@gmail.com}
}

\begin{document}
\maketitle

\begin{abstract}
 In this work, we introduce \textbf{AX}olotl, a self-study aid designed to guide students through the basics of formal reasoning and term manipulation. Unlike most of the existing study aids for formal reasoning, \textbf{AX}olotl is an Android-based application with a simple touch-based interface. Part of the design goal was to minimize the possibility of user errors which distract from the learning process. Such as typos or inconsistent application of the provided rules. The system includes a zoomable proof viewer which displays the progress made so far and allows for storage of the completed proofs as a JPEG or \LaTeX\ file. The software is available on the google play store and comes with a small library of problems. Additional problems may be opened in \textbf{AX}olotl using a simple input language. Currently, \textbf{AX}olotl supports problems that can be solved using rules which transform a single expression into a set of expressions. This covers educational scenarios found in our first-semester introduction to logic course and helps bridge the gap between propositional and first-order reasoning. Future developments will include rewrite rules which take a set of expressions and return a set of expressions, as well as a quantified first-order extension.
\end{abstract}

\section{Introduction}
Logic has, over the past century, moved from an esoteric subject studied and used, in its abstract form, by the few, to a subject pervasive in the modern world. This pervasiveness is mostly due to the ubiquity of computing technology within modern society, the foundations of which rest in the realm of mathematical logic. With this in mind, one would expect formal logic to encompass a significant portion of undergraduate computer science education, however, this is unfortunately not the case~\cite{Makowsky2017}. Part of this problem seems to be the tremendous gap between logic studied in the abstract form and its application within computing technology. 

While many can see the importance of understanding {\em Boolean algebra} when one wants to write a correct \textit{if statement}, the formal theory of propositional logic seems far removed from issues like program failure or buggy software, yet it is precisely in these situations where it has helped through verification~\cite{Silva2008} and model checking~\cite{Visser2003} techniques. The software we introduce in this work does not directly provide a way to close this gap but rather focuses on getting students comfortable with formal systems as early as possible during their undergraduate education. While our work is not the first system, designed during the last few decades, with this goal in mind, to the best of our knowledge, our system is one of the first to approach self-study of formal systems using mobile technology and without limiting the system to a particular formalism, nor limiting its scope. While there exist systems such as COQ~\cite{COQ} and Isabelle~\cite{Nipkow02}, neither is particularly well suited for beginners nor can be made into a quick and easy self-study software without significant modification. In general, such formal proof systems tend to have an extensive learning curve and are thus much better suited for advanced courses.

Our focus group consists of first and possibly second-year university students who have little or no experience with formal reasoning. Systems such as the ones mentioned in the previous paragraph are beyond their scope without significantly restricting the input language and available resources~\cite{fFBK17}. At the  Johannes Kepler University (JKU), first-year students in computer science are required to take a course on logic and formal reasoning. Our software was developed to deal with educational scenarios arising from our experience over the last few years. A recent survey of the introductory logic course at the JKU~\cite{Survey} gave us insight into what problems the student had with the introduced material as well as what motivated them and increased their understanding. Furthermore, the results of this survey together with the development of the app lead to the introduction of a laboratory assignment during the winter semester 2019 iteration of the course which specifically uses \textbf{AX}olotl. Analysis of the effect of the application on student understanding and learning is planned for the near future.

The rest of the paper is as follows: In Section~\ref{related}, we discuss existing literature and educational tools.  In Section~\ref{rules}, we discuss rule-based formal reasoning in \textbf{AX}olotl. In Section~\ref{Soft},  we introduce \textbf{AX}olotl, give a brief tutorial and discuss the educational scenarios it is designed to address. In Section \ref{AXPS}, we discuss details of the implementation such as the input language and the type of problems definable in the current release. In Section \ref{future}, we discuss future work and planned extensions of the current application.

\section{Related Work}  
\label{related}
Self-study assistants and educational software for logic can be separated into three categories based on the intended user interface: computer-based, web-based, and mobile-device-based. We will, for the most part, ignore physical games such as WFF'N~\cite{wffn1966}. The majority of software discovered during our literature search was either computer-based or web-based. For the most part, those tools designed for web-based interaction did not have a mobile device friendly web-design, i.e. a variant of a computer-based system. Given that the rise of mobile technology has occurred relatively recently and that the limitations of a mobile interface do not provide the optimal environment for a traditional approach to logic, we understand this dichotomy. These limitations are something we address in Section~\ref{Soft}.

Most of the logic educational software we present below was either presented at or discussed at the following two venues: Tools for Teaching Logic (TTL)~\cite{DBLP:conf/ticttl/2011} and ThEdu~\cite{DBLP:journals/corr/abs-1803-00722}. In particular, THedu focuses on the use of theorem provers as educational tools, while TTL is much broader in scope. An early survey covering the existing logic education tools was published at TTL in 2011~\cite{Huertas:2011:TYC:2021573.2021589}. This survey outlined over 25 different existing educational tools for logic, many of which still exist today, however, none of the tools outlined in that survey were designed for mobile devices. Rather than covering the tools outlined in~\cite{Huertas:2011:TYC:2021573.2021589}, we will instead address trends present in the software therein discussed. For instance, the majority of the approaches taken by early developers focused on the development of software for a particular formal language~\cite{DanDC,10.1007/978-3-642-21350-2,Pandora}. While this is understandably a good approach within a traditional course setting, the variety of applications of logic seen today require a more flexible system, which gives the instructor the ability to freely expand and contract the formal language to a particular problem. A few existing systems taking a step in this direction are the risc program navigator~\cite{DBLP:journals/corr/abs-1202-4834}, RISCAL~\cite{DBLP:journals/corr/abs-1904-00620}, and Theorema~\cite{JFR4568}. The first two systems consider first-order logic (FOL) over finite models, while Theorema, defined in Mathematica, considers FOL over arbitrary models and using Mathematica's interface restricts the student's view of the current state of the software. Nonetheless, none of these tools has a design adequate for mobile-based use cases. 

Concerning purely mobile-based software, i.e. tools developed for Android or IOS, there is much less variety in design than what can be found for other user interface categories. We conjecture that this may have to do with the steep learning curve required to write mobile applications.  Furthermore, the existing mobile-based tools behave like calculators or focus on restricted fragments of the chosen formal system. These fragments are usually too weak or spurious to be of major use as a university-level educational aid. A good example of such a self-study aid is the mobile phone app \textit{Quantifiers!}~\cite{spoon2018}. While the system is similar to our work in that it takes a fragment of the formal system and attempts to gamify it in an educationally useful way, the resulting questions tend to be of little value. Most of the quantifier examples presented tend to be contrived, i.e. they could be written much simpler without quantification. Furthermore, term construction is never needed, since the term language consists of a fixed set of constant symbols only. Given that term, manipulation is one of the harder problems for students it is quite unintuitive for educational software to skip it. Other examples of similar software are \textit{Emojic}~\cite{Haustein2017}, which completely abstracts logic away and focuses on image-word associations, and Lewis Carroll~\cite{Lewis2012}, which is written in Scratch~\cite{ScratchMarji} and focuses on natural language syllogisms. However, unlike the other software mentioned, Lewis Carroll does have the educational value given the role syllogisms play in philosophy education. 

One mobile application that is of similar design as AXolotl is Peanoware~\cite{Peanoware} a proof construction system for natural deduction. The application has a minimalistic interface and comes with 22 built-in problems. The main goal of the app is to construct a proof for the given formula. Interestingly, the developer was able to capture the spirit of natural deduction by allowing derivations to be built both top-down and bottom-up. Furthermore, the proof construction makes use of gestures readily available on mobile devices such as drag and drop. Unfortunately, the overall design ends up impeding one's progress rather than aiding it. To give the reader an idea of the problems we are referring to, consider that no tutorial is provided with the software. While the necessary gestures are obvious, for someone new to natural deduction making the correct choice of gestures may be difficult, furthermore it may not be clear to them why a constructed proof is complete or not. At no point is one really told what they are doing.

As one may expect from a subject with its roots in philosophy, much of the existing mobile-based software is aimed at philosophy education. Note that natural deduction is a favored proof system of the subject, possibly a motivation for Peanoware. For example, the mobile app \textit{Andor}~\cite{Andor2018} focuses on the understanding of natural language statements logically. This is one type of exercise that the interactive textbook \textit{Carnap.io}~\cite{Krouse2017} provides, though using a web-based user interface without mobile-friendly settings. Similarly, Terrance Tao developed an interactive textbook~\cite{QED2018} for understanding the logic behind mathematical theorems. Integrating both the natural language interpretation with mathematical understanding is tackled by  \textit{Lurch}, a mathematical text typesetter~\cite{Carter2014UsingTP} with an integrated prover. Though this latter software is quite out of scope, it highlights the level of development computer-based tools has reached over the past decade. 

While the above outline covers many of the outliers concerning logic education software for self-study, the majority, which we have yet to mention, focus on derivation construction and proof construction assistance. A quite important example of such derivation construction tools is the \textit{Sequent Calculus Trainer} ~\cite{Ehle2017} which provides a user-friendly interface for the construction of sequent calculus proofs as well as a hint engine powered by the Z3 SMT solver~\cite{moura2008}. Note that one must use the standard sequent calculus inference rules, thus limiting the freedom of the instructor to develop more appropriate inference sets. For example, the basic logic course at the JKU uses a version of the sequent calculus better suited for an introduction to formal reasoning, i.e. rather than using sequent rules which are completely decompositional, we introduce students to rules with precise logical meaning such as \textit{Modus Tollens}. 

The already mentioned interactive textbook carnap.io~\cite{Krouse2017} also includes a proof construction interface, but for natural deduction only. However, unlike the Sequent Calculus Trainer which has buttons for each rule carnap.io requires free-form text input from the user following a particular style of natural deduction proof representation (i.e. Fitch style, Montague style, etc.). This leaves more room for errors on the student's part, furthermore, such proof representations are not appropriate outside of philosophy education, thus limiting the applicability of the system. It is understandable why proof presentations such as Fitch style, would be used instead of the tree representation typical in computer science. Drawing a large proof tree by hand is quite a taxing endeavor. Nonetheless, there is a mismatch between these formulations of proof construction and how logic can enter the typical computer science curricula and thus making the integration of these educational tools more difficult. 

Some other worthwhile mentions are the mobile app \textit{Natural Deduction}~\cite{Jukka2018} (This is a different app then Peanoware), \textit{NaDeA: Natural Deduction assistant}~\cite{Villadsen2018}, \textit{ The Incredible Proof Machine}~\cite{Breitner2016}, and \textit{SPA: Students' Proof Assistant}~\cite{Schlichtkrull2018}. Unlike most other mobile apps, Natural Deduction provides a natural deduction fitch style proof system as well as a theorem prover. This is, to the best of our knowledge, the most highly developed educational app for logic self-study. There exist mobile logic assistants which include theorem proving technology, for example \textit{Logic++}~\cite{logic2016}, but they tend to put less focus on the assisting aspect and more on the proving. NaDeA is of similar design as Natural Deduction in that the student can construct proofs and have the system prove the statement for them. However, NaDeA can also provide hints to the students, essentially guiding them through the proof. The system is implemented on top of Isabelle~\cite{Nipkow02} and thus benefits from its further development and expressive power. Of similar design as NaDeA and also implemented in Isabelle is SPA which aids students through the process of developing a proof assistant. Note that NaDeA highlights an earlier point we made concerning systems like Isabelle. They are extremely expressive but also complex to use, thus it takes work on the side of the instructor to limit the systems for first-year students. 

The final self-study software we will discuss is the Incredible Proof Machine~\cite{Breitner2016} which is a proof construction tool with a novel interface design. Rather than constructing traditional proofs, users build circuits that match the inference rules. While this is pretty standard with respect to propositional logic, it also provides an intuitive interface for more difficult calculi such as Hilbert Systems~\cite{Monk1976} and the Lambda Calculus~\cite{DBLP:books/daglib/0032840}. Unfortunately, it is once again a web-based logic education tool without support for mobile devices. 

Concerning our contribution to the plethora of existing software, \textbf{AX}olotl, we provide a mobile-based software, with an intuitive interface, and an expressive formal language for specifying inference rules. Essentially, the user is provided with a list of goals that they must empty using the provided inference rules. After inference application, the proof view updates so students can see their progress. So far the system can handle any variation of the standard Hilbert, sequent, and natural deduction calculi as well as encodings of the resolution and tableaux calculus within sequents. Furthermore, basic equational reasoning can be performed. However, inference rules can only have a single primary formula in their conclusion and any number of auxiliary formula. In some casesm terms need to be constructed, especially for Hilbert and natural deduction calculi. When this is required, a calculator-like window opens providing an interface similar to other mobile-based applications like the Sequent Calculus Trainer~\cite{Ehle2017}, however, at no point does the student have to type the term, thus avoiding typographical errors common to other systems.

\section{Rule Base Logical Reasoning}
\label{rules}
The scope of the tool lies between propositional logic and FOL, one can think of it as the quantifier-free fragment of FOL, or, equivalently as the extension of propositional logic where every atomic formula is an atomic predicate applied to terms consisting of constants and function symbols. By using the tool students shall learn something beyond a typical introduction to propositional logic, it provides preparation for reasoning in FOL. In particular, our implementation of natural deduction and Hilbert systems presents some of the less intuitive rules for beginners as a type of variable instantiation. We cover this in more detail in later sections of the paper. Before continuing, we would like to mention that by  natural deduction and Hilbert systems we are not limiting our selves to particular instances of these techniques discussed below, but mainly the techniques associated with these calculi. 

The logical core may be thought of as follows: we have a proof ``situation'' consisting of a set of ``expressions'' $E_1,\cdots ,E_n$ where each expression is an arbitrarily nested application of ``function'' symbols to ``constants''. In the initial proof situation, there is typically only one such expression. We have inference rules of form $\Delta,\cdots \Rightarrow \Delta, \cdots$, where $\Delta$ denotes an arbitrary set of expressions and $\cdots$ denotes an expression ``pattern'' that may contain ``variables''. Note that the variables on the left and right side of $\Rightarrow$ need not match, i.e. new variables may be introduced. On the left-side only a single pattern may occur, on the right side, there may be any number of patterns. A rule with zero patterns on the right side is an ``axiom''.

An inference rule can be ``applied'' to a proof situation if the ``$\cdots$'' on the left side of the rule matches (by substituting the variables by the sub-expressions) one of the expressions in the situation; the application then replaces the expression by the ``$\cdots$'' on the right side of the rule. Here in the ``$\cdots$'' pattern, all variables which also occur on the left side of the rule are replaced by the expressions determined by the matching of the left side; for all other variables arbitrary new expressions (e.g., chosen by the human) may be substituted. Thus, by application of an axiom, expressions are removed. The proof is complete when the situation is ``empty'', no expressions left.

The didactic goal is to teach students about the basic concepts of ``rules'', ``patterns'', and ``matching'', and furthermore  the appropriate selection and application of rules matching given situations, such that by a certain number of selections and applications a certain goal is achieved.

We have experienced in our course that this
thinking process is the major stumbling block for understanding
the process of ``first-order'' reasoning; this process is
independent of the particular inference rules of FOL (and other formal calculi as well). In particular, we have noted that student performance drops when these abstract notions are introduced. 

In particular, one may use this framework to demonstrate
``sequent style'' reasoning (as done in our course) by
having every expression encode a sequent; the inference
rules are then (encodings of the) inference rules
of the sequent calculus.

The next section demonstrates by a concrete example of how
this reasoning process is implemented in the \textbf{AX}olotl software.
\section{Using \textbf{AX}olotl}
\label{Soft}
The current release of \textbf{AX}olotl requires at least API 23 of the Android operating system. This is equivalent to having release 6.0 or Marshmallow. The application is relatively small requiring only 12 megabytes of space. In order to save pictures of completed proofs, the application requires access to the user's internal storage directory. The application is available in the google play store\footnote{\href{https://play.google.com/store/apps/dev?id=6871709124320468307}{https://play.google.com/store/apps/dev?id=6871709124320468307}}.

\begin{figure}
\begin{center}
\fbox{\includegraphics[scale=\horizontalsize]{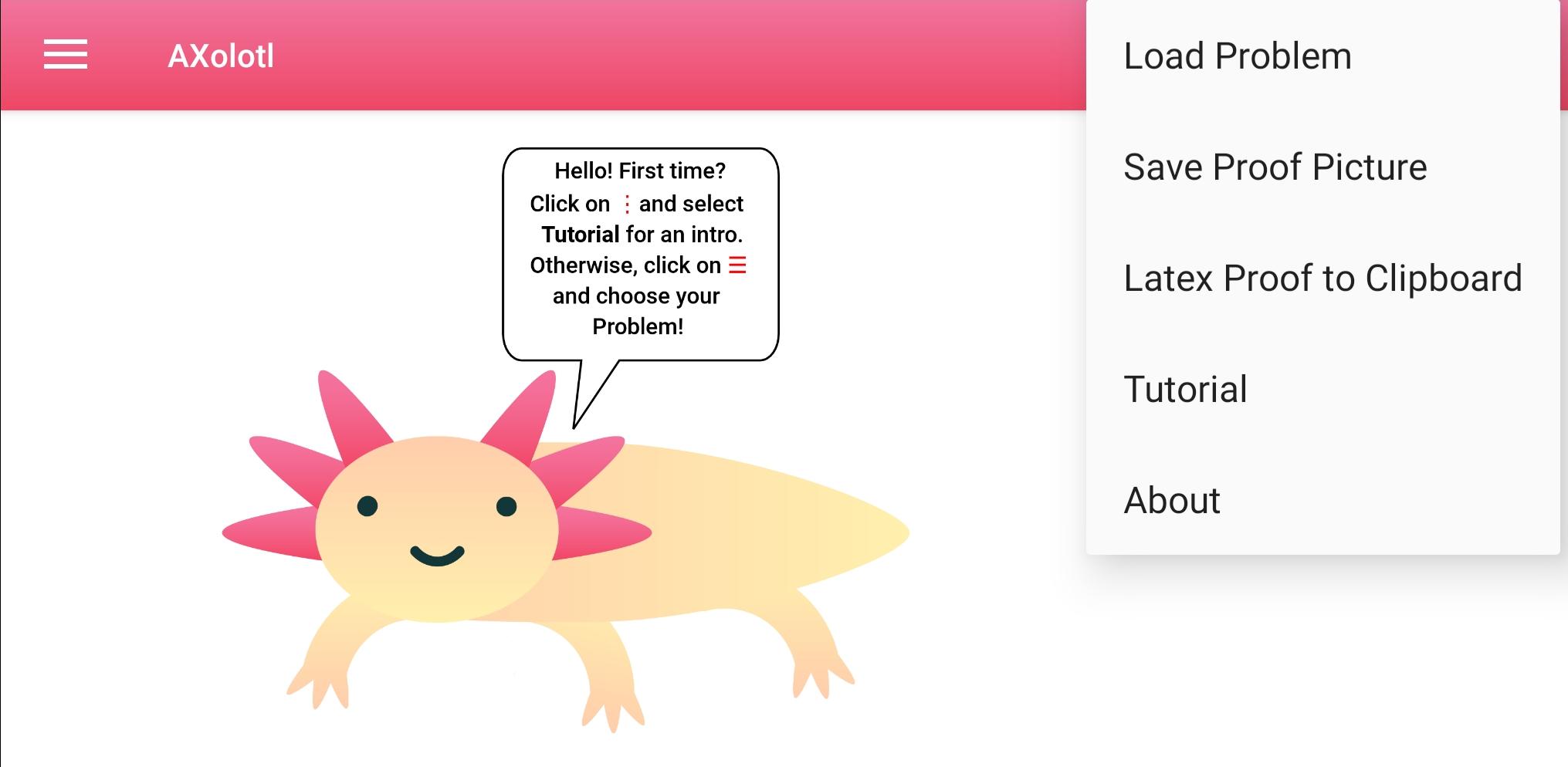}}
\end{center}\caption{The options menu of \textbf{AX}olotl.}
\label{image2}
\end{figure}

When \textbf{AX}olotl is first opened (see Figure~\ref{image2}) our axolotl mascot appears providing hints concerning what to do next. Tapping on the ``kebap'' button opens the options menu displayed in Figure~\ref{image2}. If one would like to open an \textbf{AX}olotl file which contains a problem not found in the library then the ``Load Problem'' option may be used.

The following two options concern saving completed proofs: one may either save the proof as a JPEG image which may be found in the gallery of the device in the \textbf{AX}olotl directory or export it as a \LaTeX\ file. The file is saved to the clipboard and may be copied to any other software available on the device. 

\begin{figure}
\begin{center}
\fbox{\includegraphics[scale=\verticalsize]{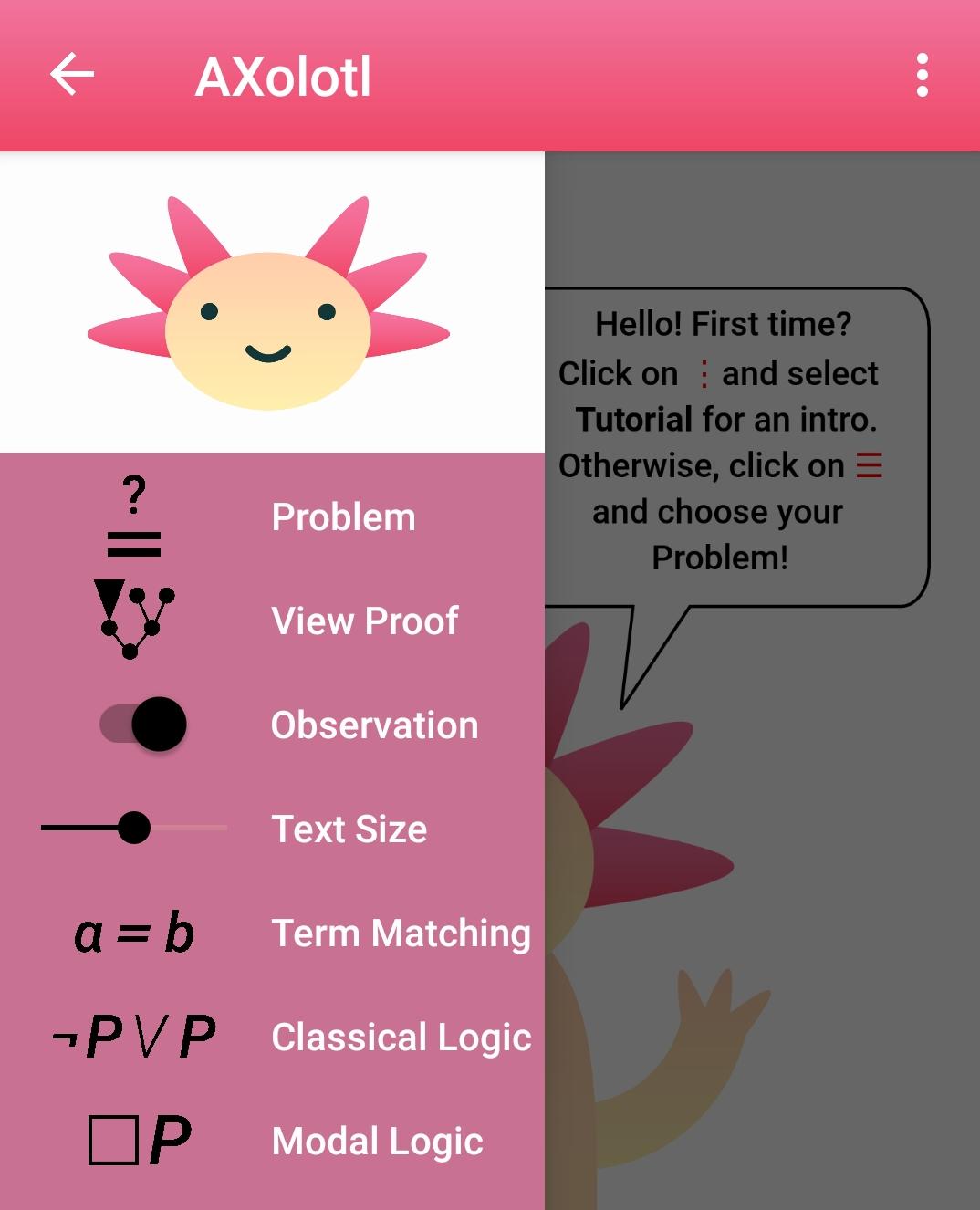}}
\end{center}\caption{The drawer menu of \textbf{AX}olotl.}
\label{image3}
\end{figure}

\begin{figure}
\begin{center}
\fbox{\includegraphics[scale=\horizontalsize]{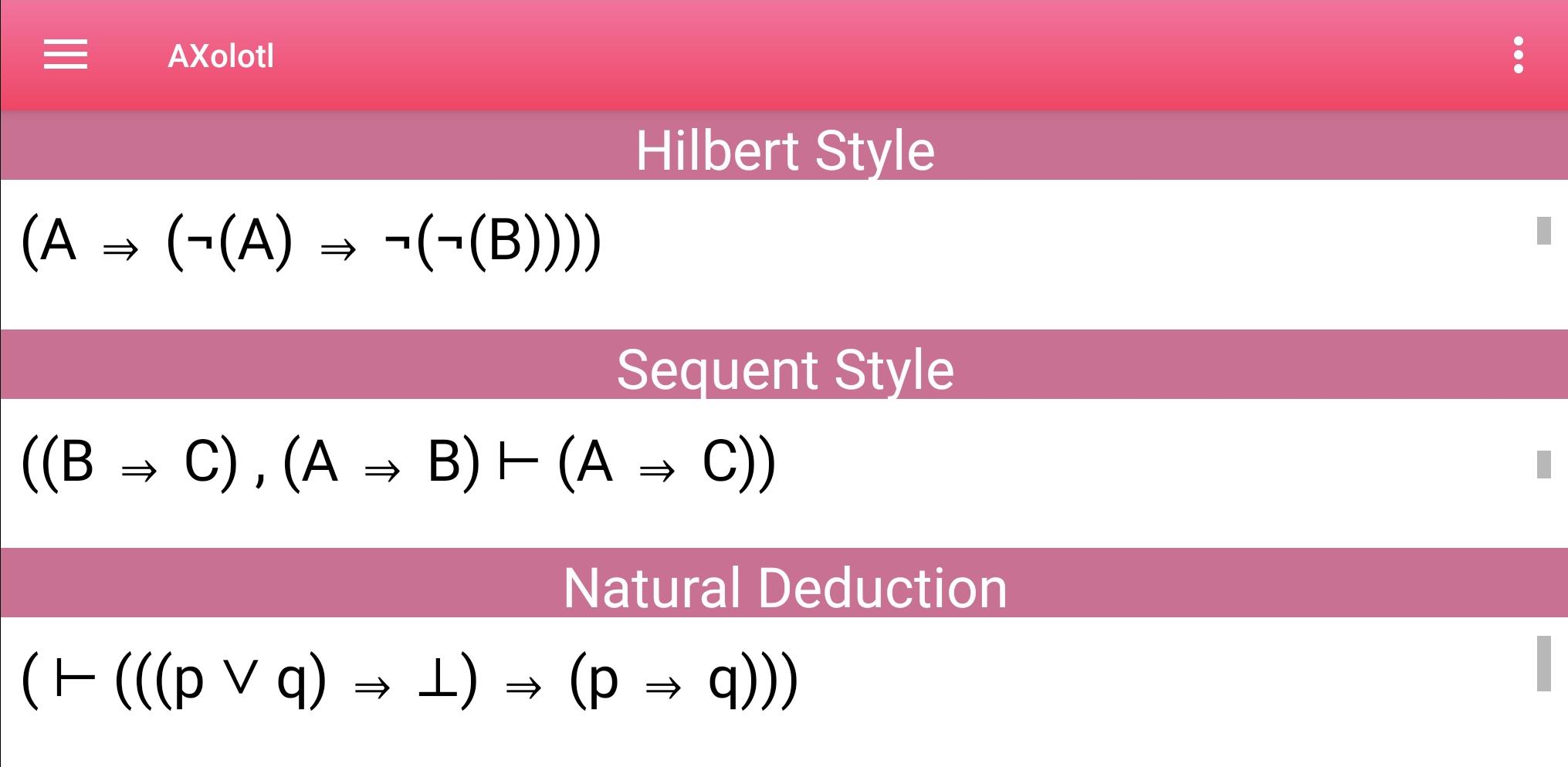}}
\end{center}\caption{Problem library for classical logic problems.}
\label{image4}
\end{figure}

The ``Tutorial'' option concerns first time users, it provides a short tutorial covering how to use the system. When the tutorial ends, or if one taps on the about button, information concerning the developers is provided. 

If instead of opening the ``Options'' menu one taps on the ``hamburger'' button, the drawer menu opens as seen in Figure~\ref{image3}. The first two buttons of the drawer menu concern the currently loaded problem, of which there currently is none. 

\noindent Concerning ``Observation'', \textbf{AX}olotl provides users with a step-by-step application of inference rules that demonstrates how an expression is changed by the chosen inference rule. 

\noindent This can be quite verbose and can slow down the pace of the user. When a user is advanced enough ``Observation'' can be turned off providing a much smoother and quicker experience.  

Concerning ``Text Size'', when expressions get large, using the standard font size may result in much-unwanted scrolling. To avoid this, we allow the user to adjust the font size to be between 10 and 50 SP (Android's unit for screen invariant font size). 
The rest of the menus concern the built-in libraries of problems. Future upgrades of the system will contain extensions of these libraries.

Let use now consider the library for classical logic, see Figure~\ref{image4}. Notice that there are three categories of problems, namely Hilbert style, sequent style, and natural deduction. As one might guess, these categories denote the style of deduction used to solve the problems. As was done in the tutorial available within the app, we will showcase the software using sequent style deduction and transitivity of implication (the problem visible in Figure~\ref{image4}). A proof of this property would be written as follows using Gentzen's sequent calculus:

\begin{prooftree}
\AxiomC{$A \vdash A$}
\RightLabel{$weak$}
\UnaryInfC{$(B\rightarrow C),A \vdash A, C$}
\AxiomC{$ C \vdash  C$}
\RightLabel{$weak$}
\UnaryInfC{$ B,C,A \vdash  C$}
\AxiomC{$ B\vdash B$}
\RightLabel{$weak$}
\UnaryInfC{$ B,A\vdash B, C$}
\RightLabel{$\rightarrow:l$}
\BinaryInfC{$ B,(B\rightarrow C),A \vdash  C$}
\RightLabel{$\rightarrow:l$}
\BinaryInfC{$(A\rightarrow B),(B\rightarrow C),A \vdash  C$}
\RightLabel{$\rightarrow:r$}
\UnaryInfC{$(A\rightarrow B),(B\rightarrow C) \vdash   (A\rightarrow C)$}
\end{prooftree}

\begin{figure}
\begin{center}
\fbox{\includegraphics[scale=\horizontalsize]{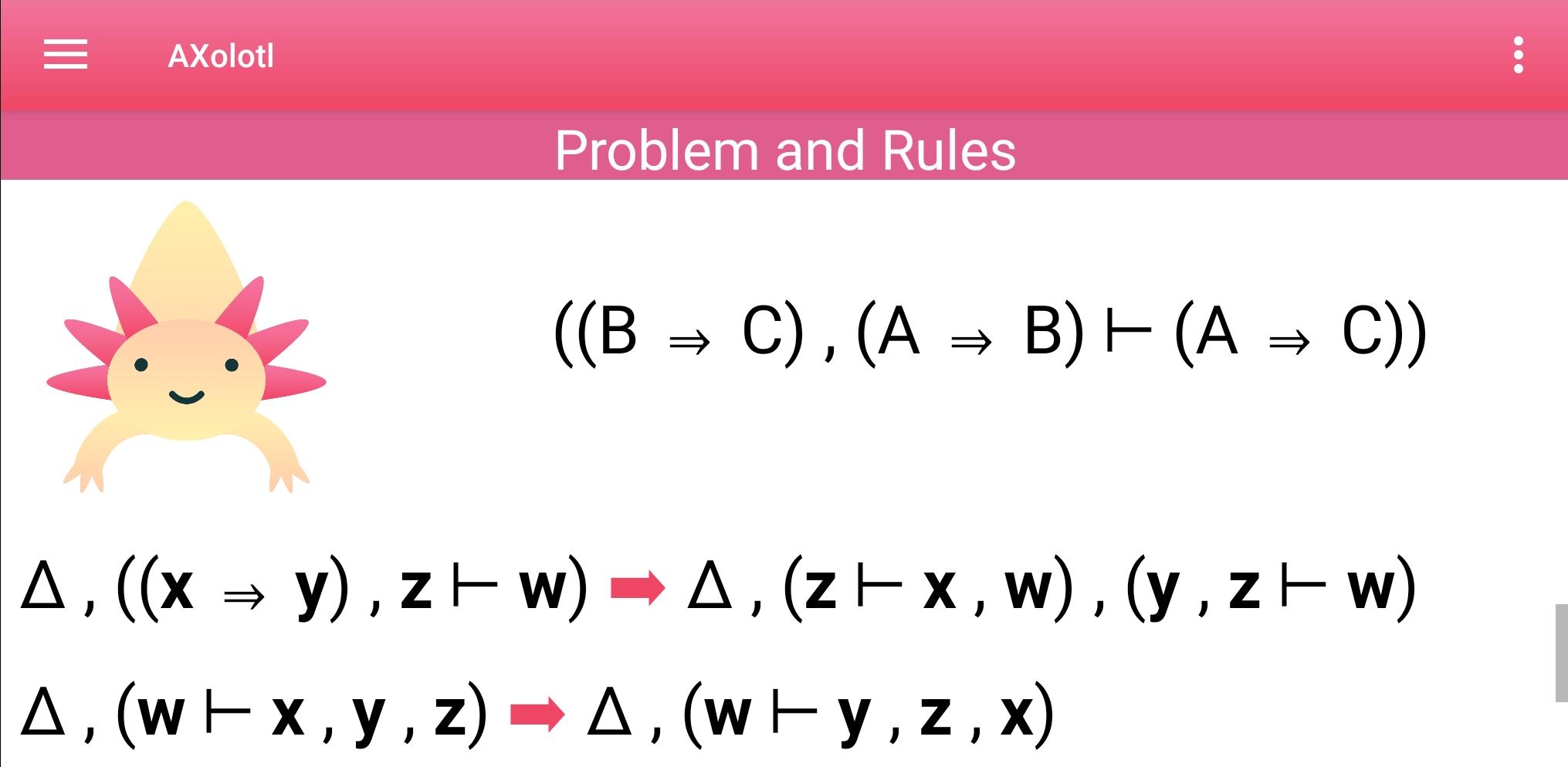}}
\end{center}\caption{Problem display.}
\label{image5}
\end{figure}

By tapping on the problem a new display will appear containing the selected problem, the rules and an image of our mascot (See Figure~\ref{image5}). Note that the rules are displayed in a succinct form which, for more complex rules, may be hard to read. Before moving on, we would like to discuss our use of $\Delta$ in the inference rules. If there is more than a single goal in the goal display, any additional
goals are abbreviated by $\Delta$. Consider $\Delta$ as an additional context that will be ignored during the application of an inference rule. 

\begin{figure}
\begin{center}
\fbox{\includegraphics[scale=\horizontalsize]{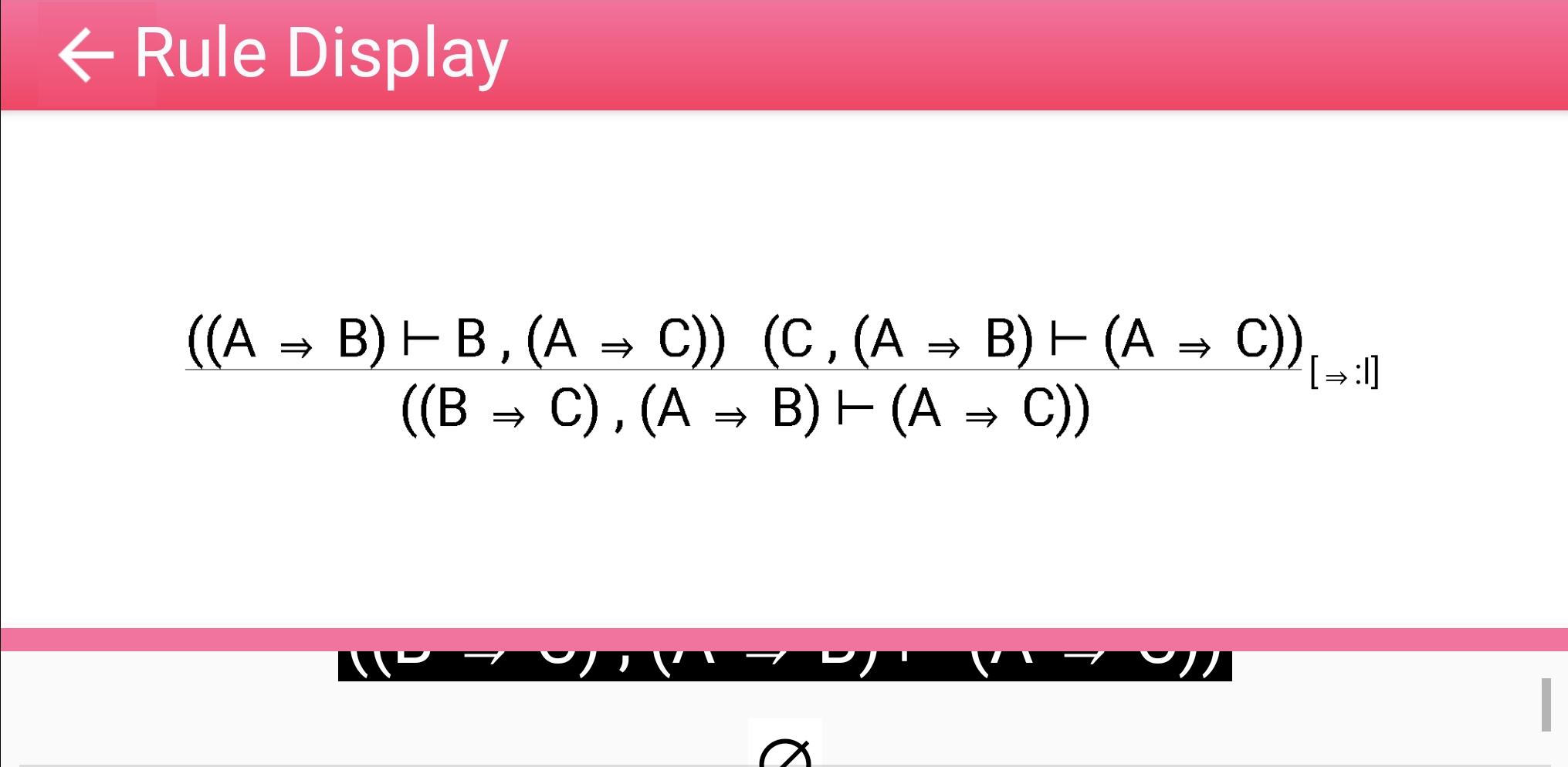}}

\end{center}\caption{Rule display.}
\label{image6}
\end{figure}

If one long taps on a rule, a new display opens which shows a pretty printed form of the rule (see Figure~\ref{image6}). Notice that the rule display allows you to select an $\emptyset$. This displays the rule without instantiating the formula variables. The rule, as displayed in Figure~\ref{image6}, is instantiated by our selected problem. 

Once a rule and a problem have been selected in the display shown in Figure~\ref{image5}, they are highlighted black, one may swipe right using a fling motion to proceed to the next screen. When \textit{observation} mode is activated \textbf{AX}olotl will present the user with the individual instantiation of the formula variables.

\begin{figure}
\begin{center}
\fbox{\includegraphics[scale=\horizontalsize]{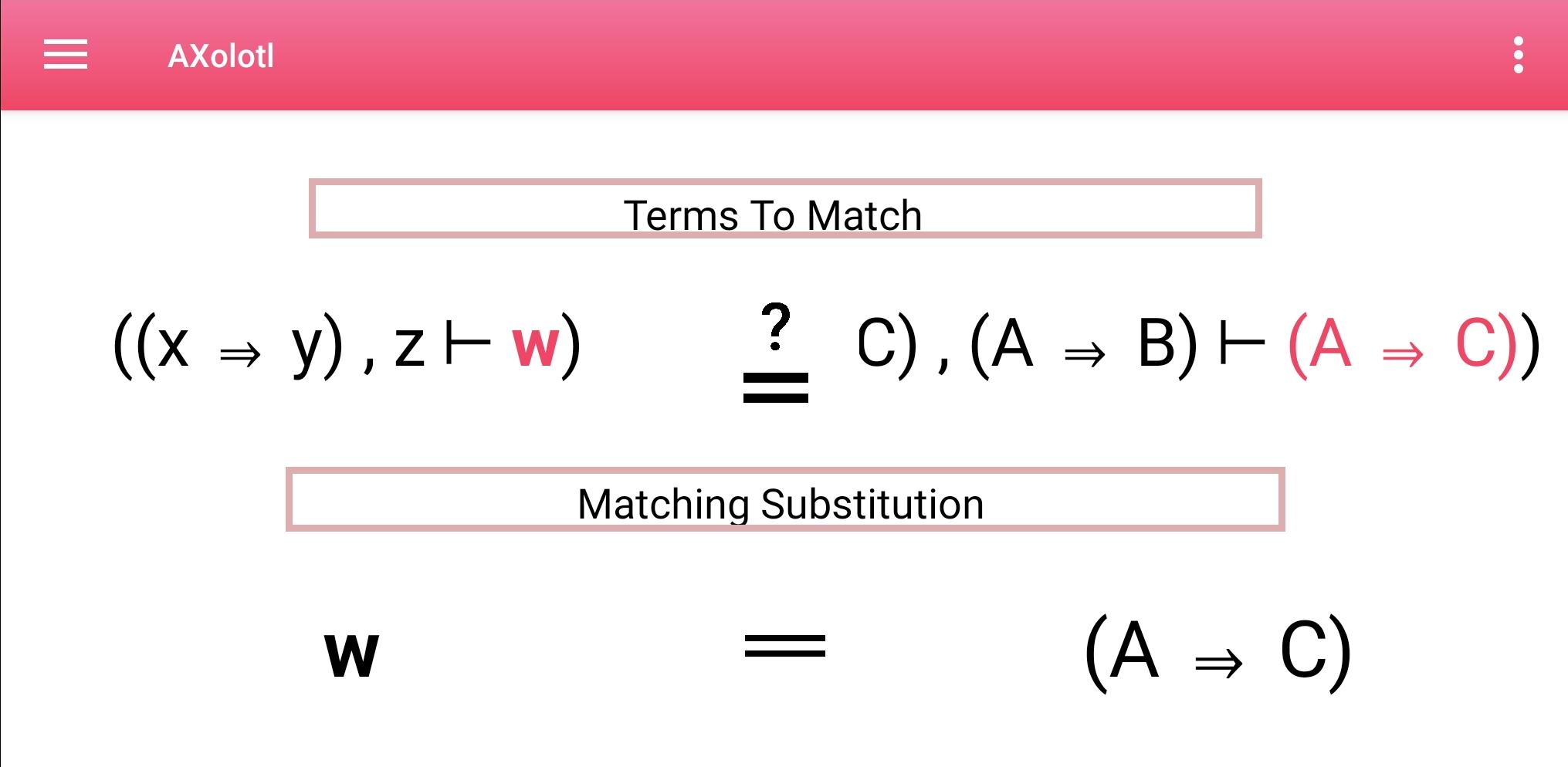}}
\end{center}\caption{Instantiation of the variable \textit{w}.}
\label{image7}
\end{figure}

\begin{figure}
\begin{center}
\fbox{\includegraphics[scale=\horizontalsize]{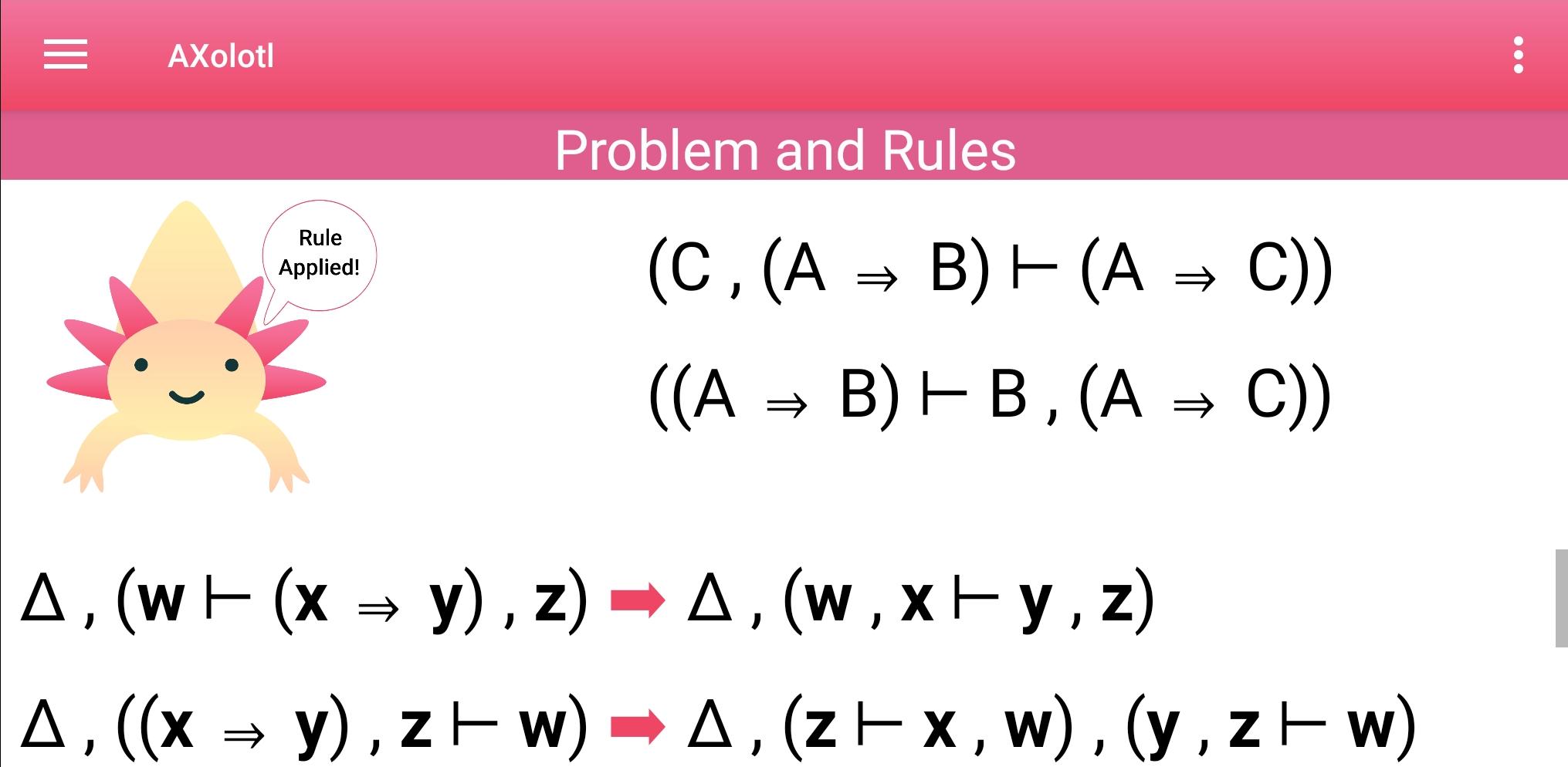}}
\end{center}\caption{Problem display after rule application.}
\label{image8}
\end{figure}

In Figure~\ref{image7}, the instantiation of the formula variable \textbf{w} is displayed.  The top portion of the display ``terms to match'' indicates which part of the rule is matched to which part of the goal.  The lower portion of the display ``matching substitution'' shows the precise substitution of \textbf{w}. Upon swiping left using a fling motion, the instantiation of the next variable will be displayed. Once all four instantiations have been displayed AXolotl switches to the screen displayed in Figure~\ref{image8}.
 
 Note that at any point one can swipe right to left with a fling motion to go back to the previous display. On the problem display screen, this motion will undo the previous rule application unless there are no rule applications to undo. 
 
\begin{figure}
\begin{center}
\fbox{\includegraphics[scale=\horizontalsize]{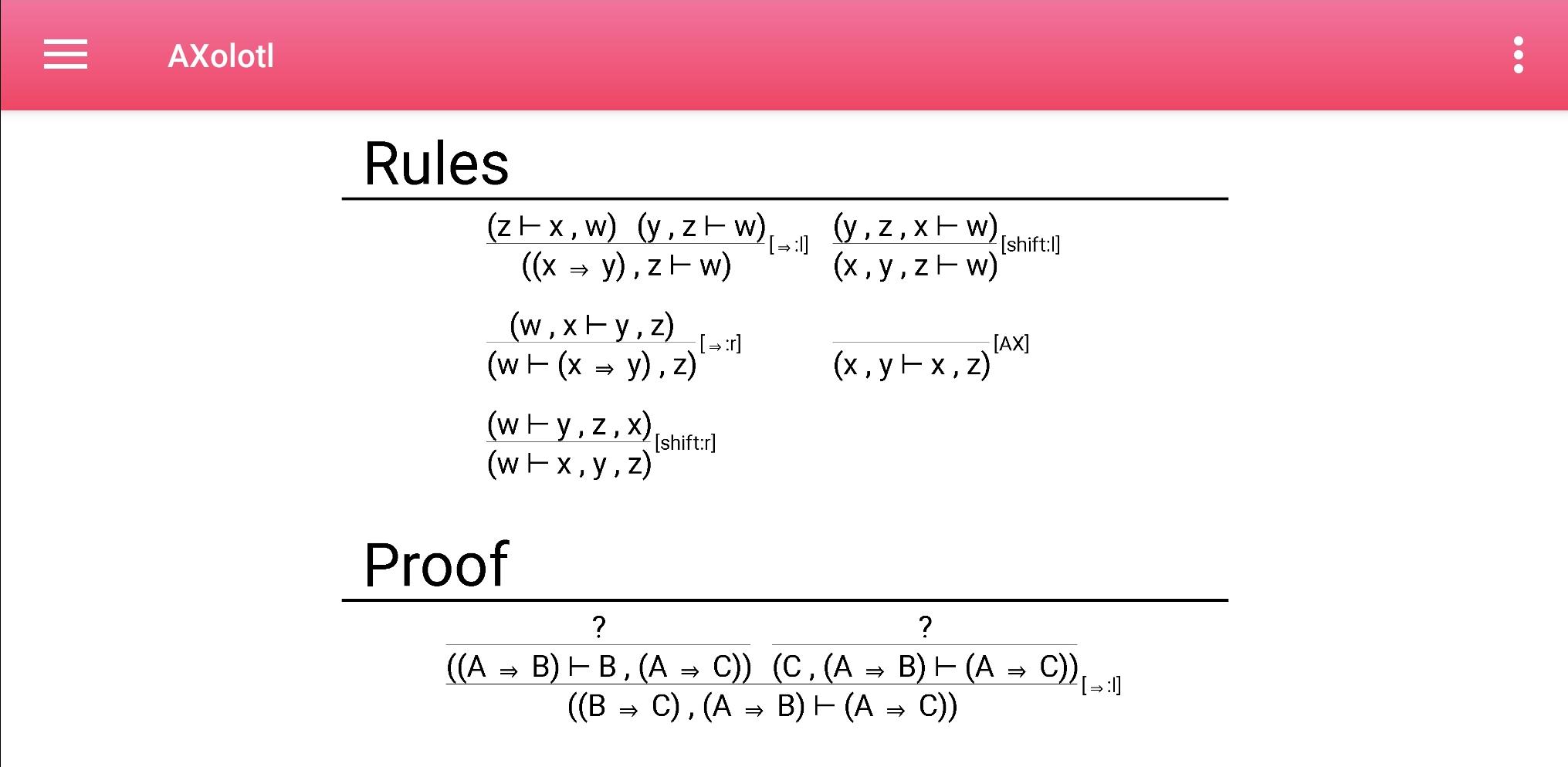}}
\end{center}\caption{The proof view after one rule application.}
\label{image9}
\end{figure}

Now if we tap on the ``hamburger'' menu and select ``View Proof'' the progress we have made so far is displayed (see Figure~\ref{image9}). The proof view is split into two parts, the rule list and the proof. The rule list contains all the allowed rules for the given problem. The proof contains the current state of the proof. Notice that the two branches end in question marks. This means that the two branches are currently open, i.e. do not end with axioms. 
 
\begin{figure}
\begin{center}
\fbox{\includegraphics[scale=\horizontalsize]{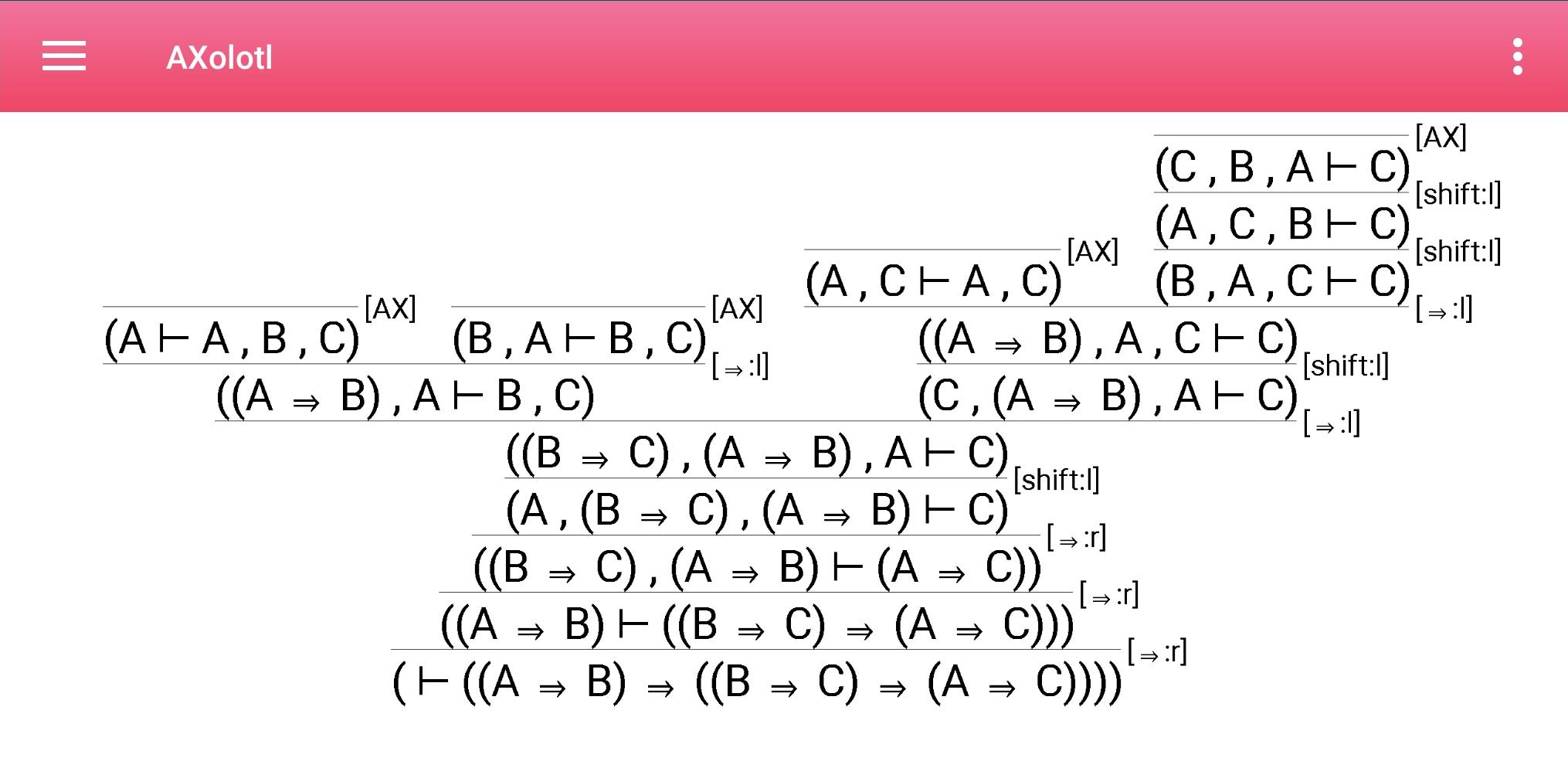}}

\end{center}\caption{The proof after the exhaustive application of inference rules to the problem.}
\label{image10}
\end{figure}

\begin{figure}
\begin{center}
\fbox{\includegraphics[scale=\horizontalsize]{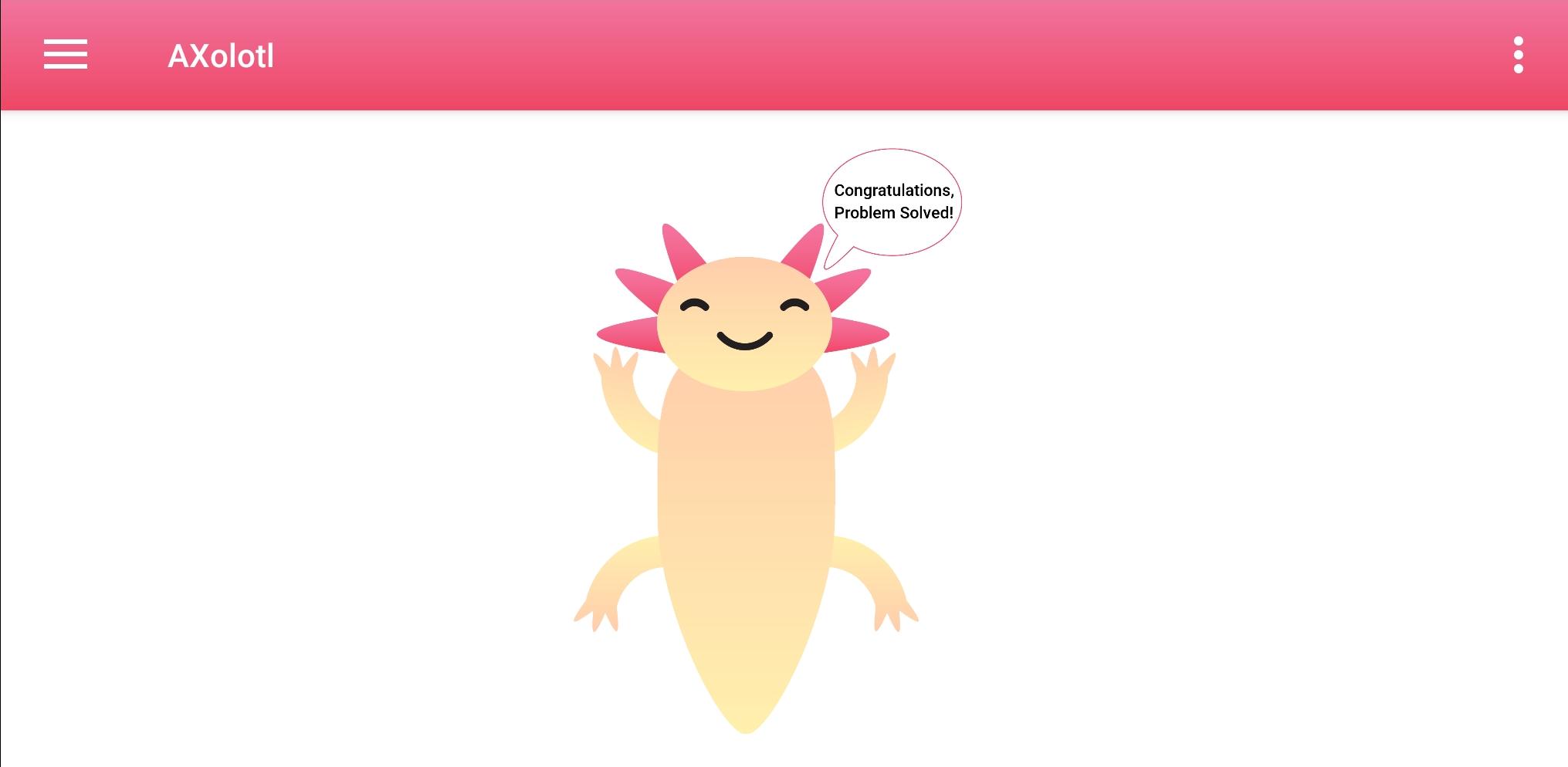}}
\end{center}\caption{Displayed when goal list is emptied.}
\label{image11}
\end{figure}

If we were to apply inference rules exhaustively the resulting proof would be as displayed in (see Figure~\ref{image10}). When an inference rule is applied which empties the goal set an animation is displayed ending with the image seen in Figure~\ref{image11}.

The proof can also be exported as a latex file with page size set to A2. The rules are not included in the latex output, only the proof. 

\begin{figure}
\begin{center}
\fbox{\includegraphics[scale=\horizontalsize]{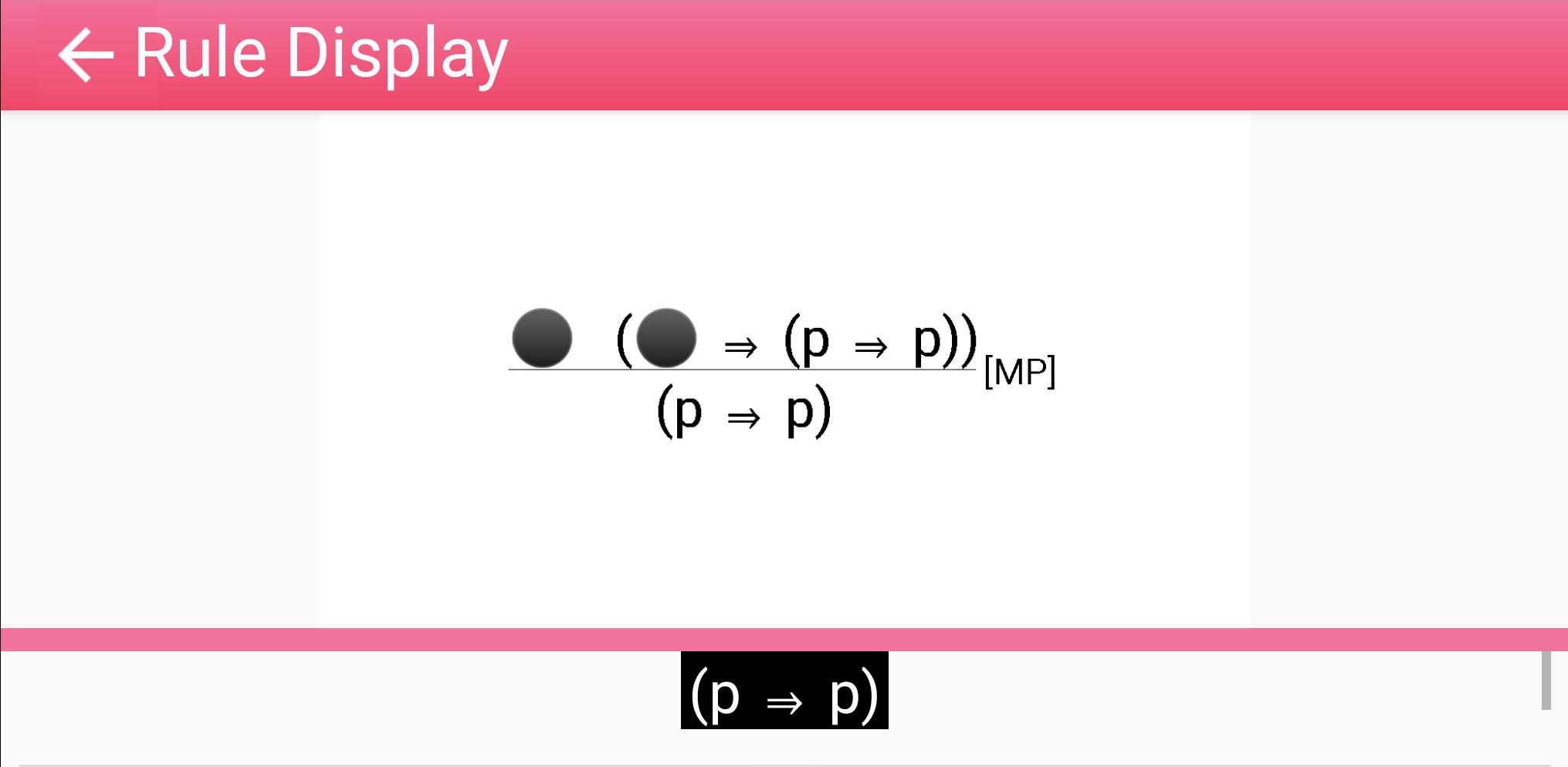}}
\end{center}\caption{A rule with a hole to be filled by the user.}
\label{image12}
\end{figure}

Note that all inference rules used in this example do not introduce new variables. This is typical for sequent calculus style proofs, however, the fundamental deduction rule of Hilbert systems breaks this assumption, that is modus ponens: 

\begin{prooftree}
\AxiomC{$x$}
\AxiomC{$x\Rightarrow y$}
\RightLabel{MP}
\BinaryInfC{$y$}
\end{prooftree} 

\noindent The conclusion of the rule does not contain the variable $x$. This means that \textbf{AX}olotl requires input from the user specifying what term is to replace $x$.

Consider proving P $\Rightarrow$ P using modus ponens and the axioms:
$ x\Rightarrow (y \Rightarrow x)$ and $(x\Rightarrow (y\Rightarrow z))\rightarrow ((x\Rightarrow y) \Rightarrow (x\Rightarrow z))$.

\begin{figure}
\begin{center}
\fbox{\includegraphics[scale=\horizontalsize]{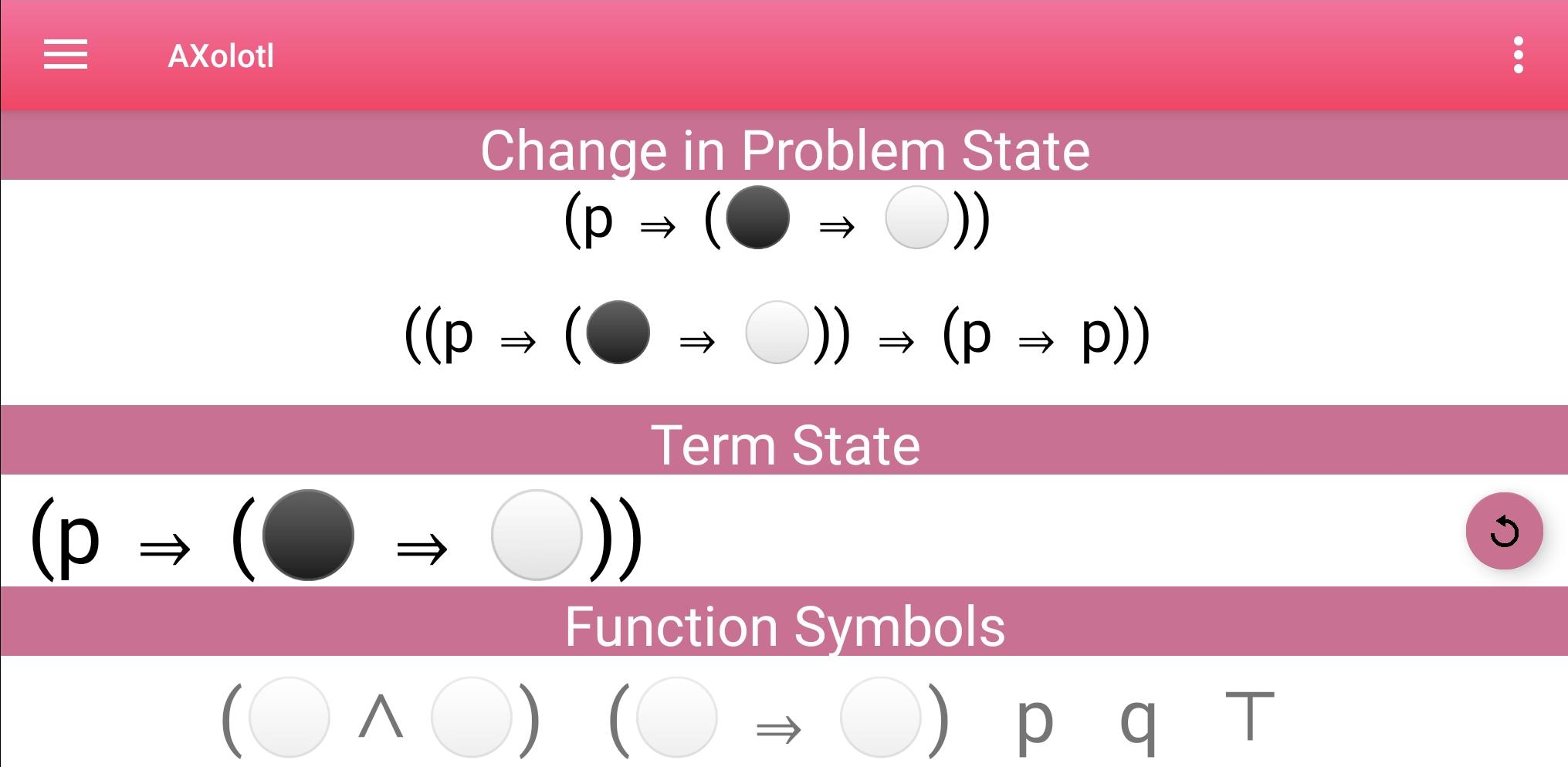}}
\end{center}
\caption{The term construct display allows the user to construct terms from the allowed function symbols.}
\label{image13}
\end{figure}

If we select the inference rule modus ponens the rule display will contain the contents of Figure~\ref{image12}. The hole visible in Figure~\ref{image12} denotes the value which must be feed to \textbf{AX}olotl. If we swipe right and attempt to apply this rule we will eventually come across the screen displayed in Figure~\ref{image13} which can be used to construct terms. 

Note that turning off ``Observation'' does not mean that Figure~\ref{image13} can be skipped, \textbf{AX}olotl does not provide instantiations of variables whose value cannot be derived from context. There are two types of holes displayed in  Figure~\ref{image13}, The selected hole and the unselected holes. There may be only one selected hole at a time but there may be any number of unselected holes. Once there are no holes of either type present one may swipe right using a fling motion. If at any point one swipes left the constructed term is erased and one must start from scratch. 

The ``Change in Problem State'' portion of the display shows what the goals list will look like based on your current instantiation of the variable. The  ``Term State'' displays the current term you have constructed, and the ``Function Symbols'' portion is a list of allowed function symbols. If a function symbol takes arguments each argument will be denoted by an unselected hole. After choosing a function symbol with holes, the left-most unselected hole will be switched to selected.  Notice that on the right side of ``Term State'' is an additional button for undoing the previous instantiation. 

\begin{figure}
\begin{center}
\fbox{\includegraphics[scale=\horizontalsize]{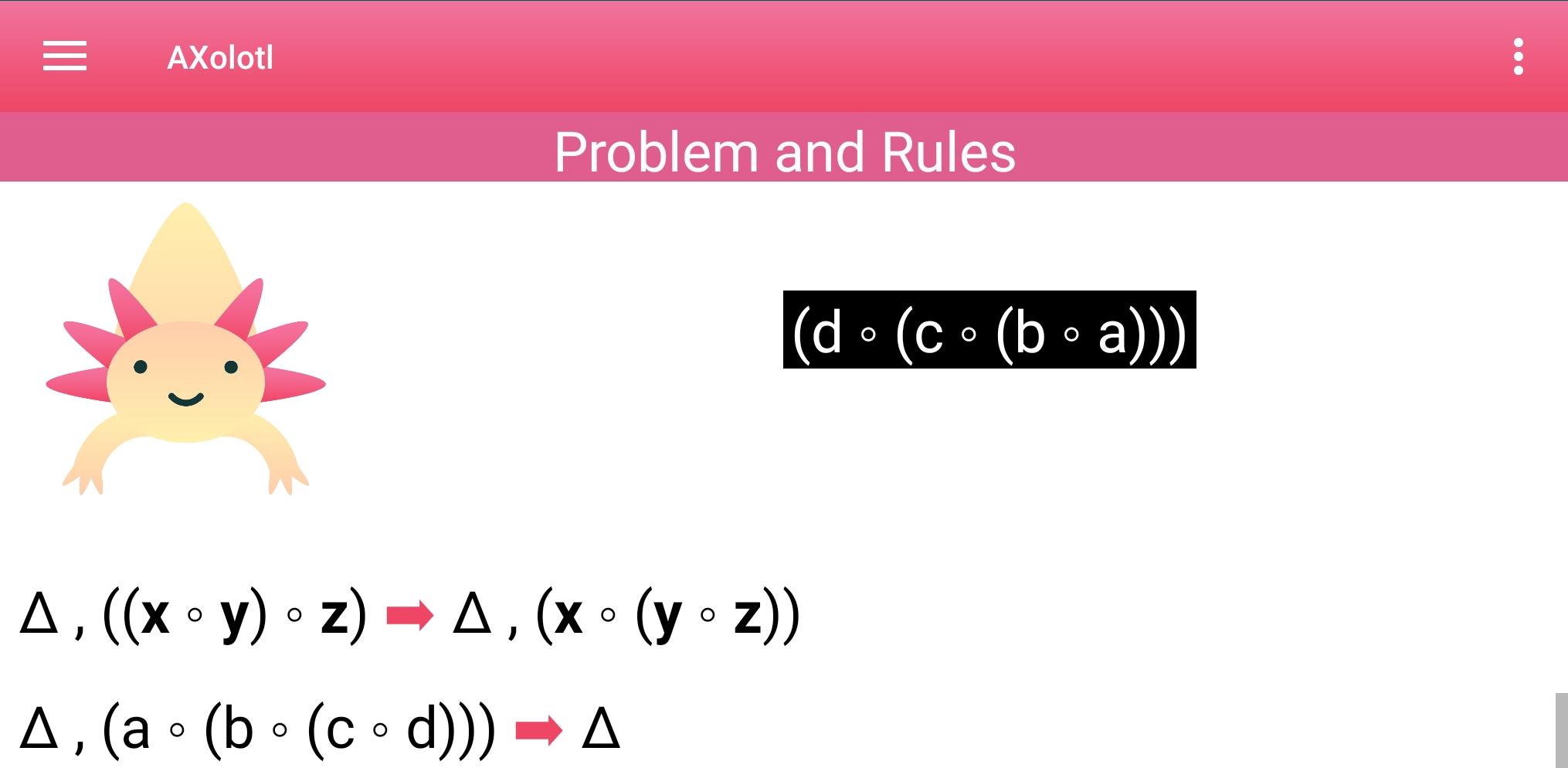}}
\end{center}
\caption{A rewriting problem in \textbf{AX}olotl.}
\label{image14}
\end{figure}

This covers the basic functionality and usage of \textbf{AX}olotl. The last thing we would like to cover in this section is how to use \textbf{AX}olotl for non-logical problems, for instance, show that two distinct permutations of a list are equivalent modulo associativity and commutativity, see Figure~\ref{image14}.

Most of the inference rules are used to transform the goal. The last inference rule is instead used to remove a particular term from the goal list. Essentially this is a variation of the equation:
 $$a \circ b \circ c \circ d \equiv d \circ c \circ b \circ a$$
 
 \noindent many other problem instances, such as linear and binary search can also be written within \textbf{AX}olotl using a similar encoding. 

\section{\textbf{AX}olotl Problem Specification}
\label{AXPS}
\begin{figure}
\begin{center}
\fbox{\includegraphics[scale=.23]{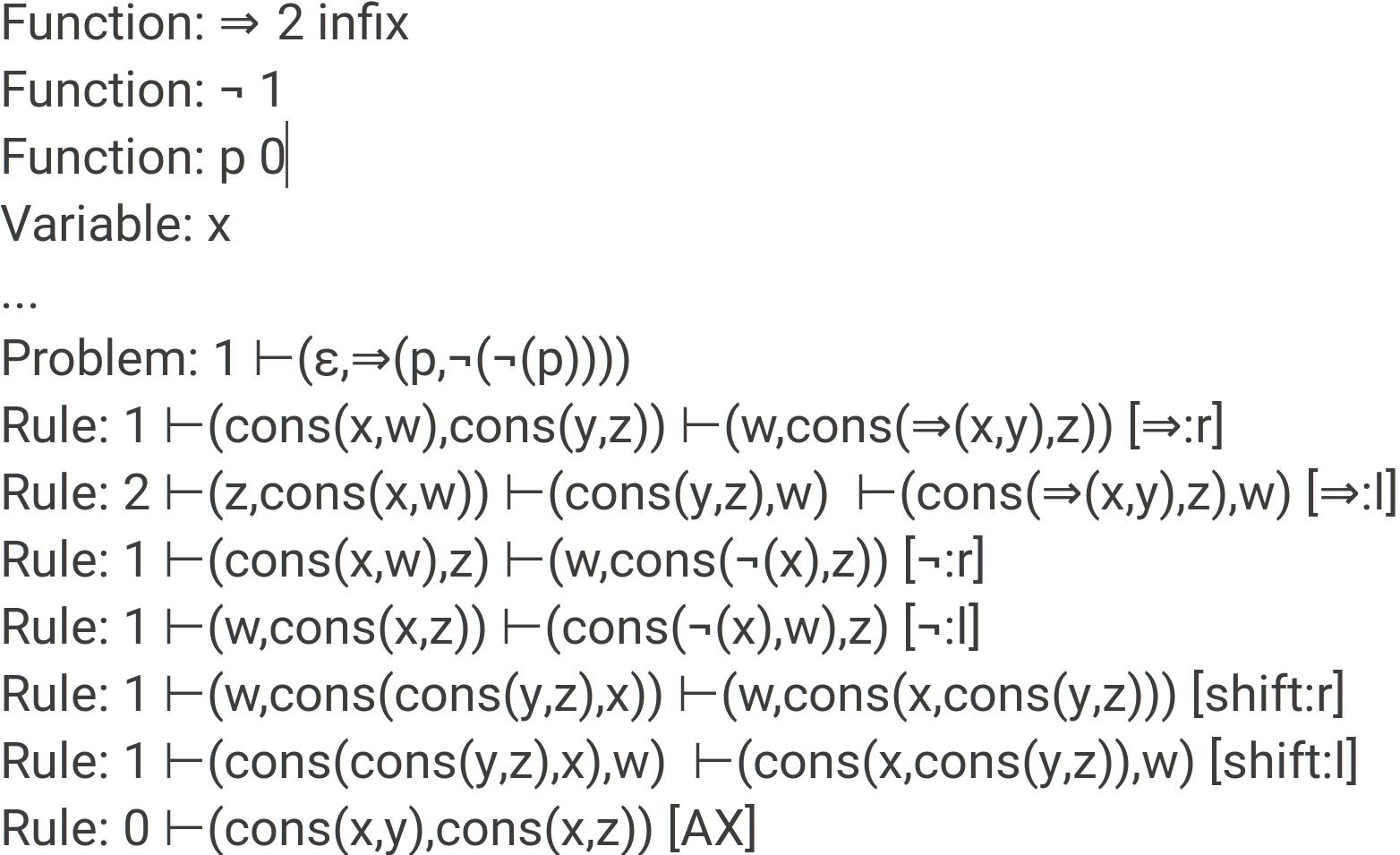}}
\end{center}\caption{A rewriting problem in \textbf{AX}olotl}
\label{image15}
\end{figure}

The problems found in the built-in library are loaded on start-up from so-called \textbf{AX}olotl files. These files are written in a simple language usable by an instructor or savvy student to design their own problems. In Figure~\ref{image15} one finds the \textbf{AX}olotl file for the first problem we discussed in Section~\ref{Soft}.

The files are read line by line and there should not be any trailing line after the final input line. Each line has a prefix which may be one of the four values: ``Function'', ``Variable'', ``Problem'', ``Rule''. All ``Function'' and ``Variable'' lines must occur before ``Problem'' and ``Rule'' lines because syntactic checks are run on the ``Problem'' and ``Rule'' statements based on the allowed symbols. There are no order restrictions between ``Variable'' and ``Function'' nor between ``Problem'' and ``Rule''. The ``Function'' lines must contain a symbol which may be alphanumeric and an arity (number of arguments). If arity is set to zero then the function is considered a constant. Each function symbol must have a unique arity. Additionally, a binary function can have infix as the last statement on the line implying that it ought to be printed infix rather than prefix. 

``Variable'' lines may only contain an alphanumeric symbol. In future releases of the software, there may be further indicators added concerning which type of variable one is defining. This would be necessary for extensions of this software to first-order where a distinction between formula variables and term variables would need to be made. 

The ``Problem'' line starts with a number indicating how many goals will be contained in the problem's goal set. In Figure~\ref{image15} there is precisely 1 goal. At this point, one may notice that the ``Problem'' line of Figure~\ref{image15} contains additional function symbols that were not defined in the file. These function symbols are predefined and have semantic meaning within the system. They are as follows: 

\begin{itemize}
\item $\vdash$ denotes sequents and must contain two lists as arguments. 
\item $cons(\cdot, \cdot)$ denotes the list constructor. 
\item $\epsilon$ denotes the empty list. 
\end{itemize}

\noindent We assume that any non-list term contained in a list cannot contain a list within it, that any singleton term $t$, which is not a variable, contained within a sequent is a list of the form $cons(t,\epsilon)$, and that the sequent symbol does not occur within a list of terms or within any term for that matter. There are a few additional assumptions concerning printing, but they do not affect the logical interpretation of the problem. 

Before we move on, the last constraint concerning the problem line is that variables cannot occur. This will most likely change in future iterations of the software. 

The ``Rule'' line starts with a number indicating the number of premises the rule has. The second rule in  Figure~\ref{image15} has two premises while the last rule has none. Rules without premises are axioms and close branches. There is no limit to the number of premises, but when outputting the proof to \LaTeX\ rules are restricted to five.

The next $n$ terms are the premises of the rule where $n$ is the number of premises. The $n+1$ term denotes the conclusion of the rule. Every rule must have a conclusion. Additionally, to the premises and conclusion, a rule may be given a name that is displayed next to it in the rule and proof display. This name comes last on the rule line and is to be written in square brackets. It is not required, but advisable.

\section{\textbf{AX}olotl in the classroom}
In the winter semester 2019, we introduced \textbf{AX}olotl into our introductory logic course as part of an optional laboratory assignment. Students usually complete the laboratory assignment when they need extra credit given their performance on previous in-class exams or when they are interested in topics that go beyond the required materials. Typically 30-40\% of the students partake in the assignment. This optional laboratory assignment is presented to the students between the modules covering propositional logic and first-order logic thus perfectly fitting the educational scenario  \textbf{AX}olotl was designed for. In particular, we used \textbf{AX}olotl as an educational aid for introducing the students to Natural Deduction. Note that Natural Deduction tends to be problematic for students being that it is easy for students to make errors in rule application and secondly, some aspects of rule application are quite close to first-order reasoning. For instance elimination of implication. Note that the Curry-Howard isomorphism between simple type theory and intuitionistic Natural Deduction maps implication introduction and elimination rules to the same $\lambda$-terms as universal introduction and elimination, see Chapter 3~\cite{Girard:1989:PT:64805}. 

\begin{figure}
\begin{center}
\fbox{\includegraphics[scale=.48]{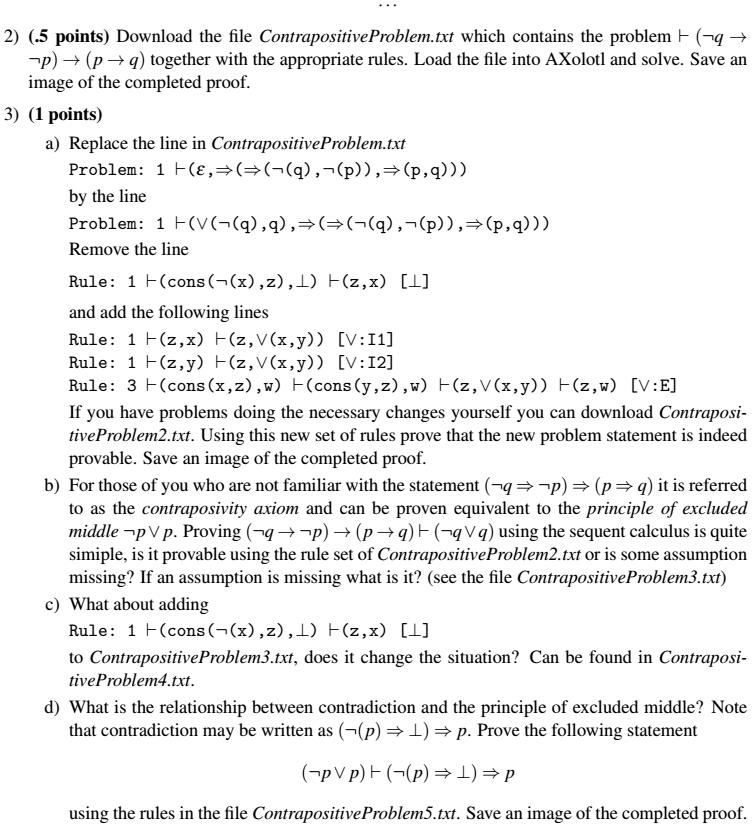}}
\caption{Excerpt from Lab assignment.}
\label{excerpt}
\end{center}

\noindent Implication within the propositional natural deduction framework can be thought of as a meta-level quantification over formula. While the Hilbert system  we use for classical propositional logic shares these properties as well given that the only inference rule, {\em Modus Ponens} is essentially implication elimination, it lacks the intuitive feeling of Natural Deduction and thus will most likely leave students frustrated. Even for the initiated, this feeling of frustration is well known and can still be felt.

\end{figure}

The goal of the introduced Laboratory assignment was to get students use to thinking in a way closer to what is required for first-order reasoning. As already mentioned in the previous paragraph, the standard Natural Deduction calculus, unlike the sequent calculus or resolution, contains inference rules which correspond to the quantification inferences of first-order logic. Though this correspondence is far from equivalence, we believe it provides enough insight into the subject to help students through what seems to be the most difficult part of the course~\cite{Survey}. While we have not, as of yet, experimentally validated the tacitly mentioned hypothesis outlined above, we are currently working on the development of experimental scenarios that can test our hypothesizes. 

Concerning the excerpt from the laboratory assignment provided in Figure~\ref{excerpt} and how it relates to the above discussion one only needs to consider how the proofs of these various statements look.

\begin{figure}
\begin{center}
\fbox{\includegraphics[scale=.1]{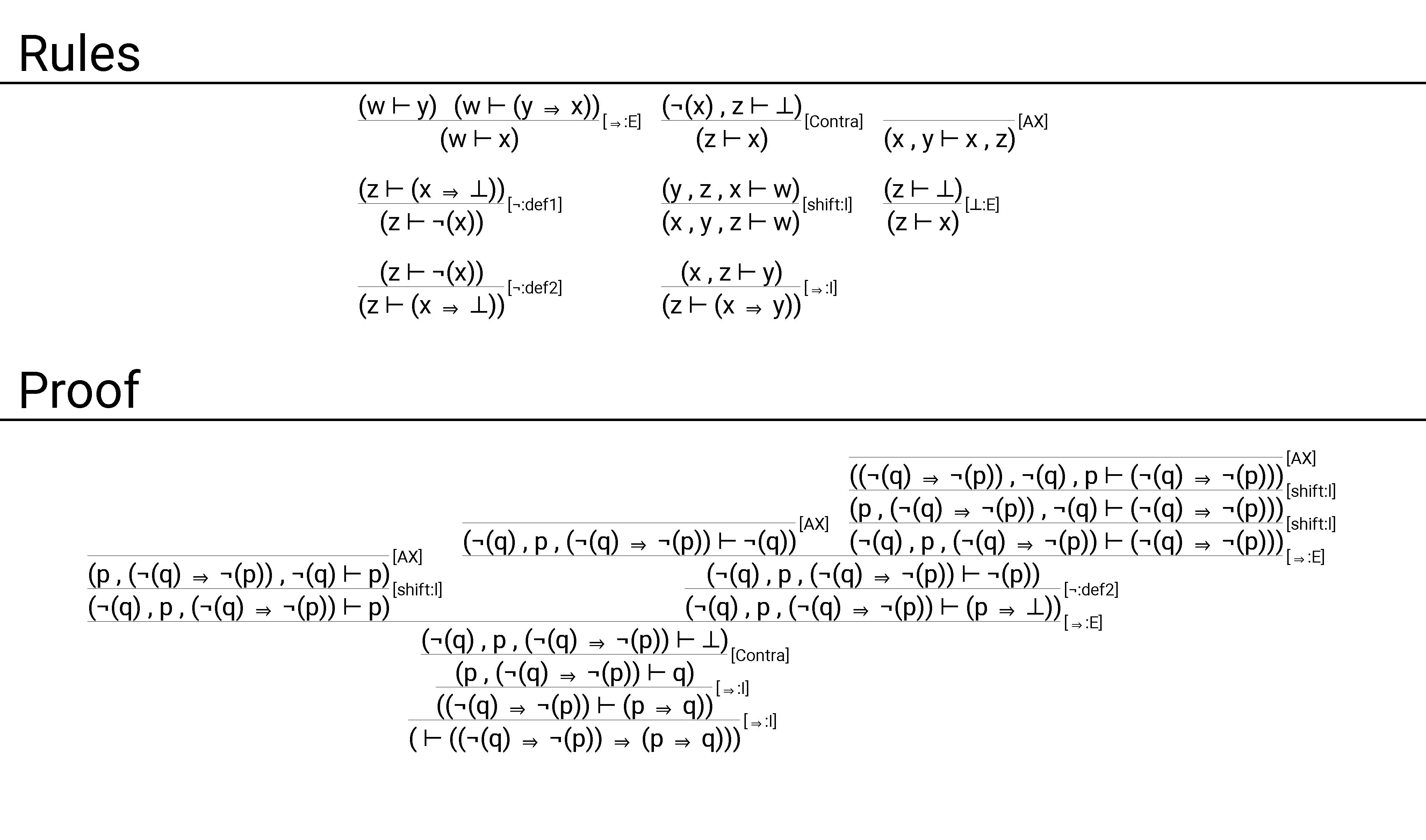}}
\caption{\textbf{AX}olotl proof of contrapositivity.}
\end{center}
\end{figure}

Notice that it not only requires the use of implication elimination but also a contradiction. The combination of these two rules provides a quite good simulation of the relationship between unification and quantification of first-order logic. While this type of exercise can be done without the aid of  \textbf{AX}olotl it would require students to be more attentive when applying inference rules something which may be hard for beginners who are prone to typographical and consistency errors. 

The rest of the questions from the Laboratory excerpt concern similar issues but using different inference rule sets and different assumptions. Given the flexibility of \textbf{AX}olotl's input language, we can easily switch between rule sets while maintaining a user-friendly environment. 

\section{Future work}
\label{future}
In previous sections, we have briefly mentioned a few points which we plan to tackle in the near future. One of the most important developments will be extending the input language as well as the system to allows rules with multiple conclusions. 

\noindent Simple and important calculi such as resolution~\cite{Leitsch:1997:RC:260906} and tableaux~\cite{DAgostino1999} cannot be expressed in a natural way using our rule descriptions; rather than the proof situation being stored as a set of expressions, as in the \textbf{AX}olotl implementation of the sequent calculus, it is stored within a single expression. 

\noindent Furthermore, our proof display is inadequate for displaying the intermediary steps within a resolution proof, this would require drawing a forest of trees. Also, this additional flexibility will allow users to expression even more calculi than the current input language allows. 

A simpler yet more challenging change would be the addition of quantifiers. This would require minimal changes to the look and feel and input language but would require extensive reworking of the current unification and matching algorithms~\cite{Baader:1998:TR:280474}. For those who are familiar with higher-order reasoning, we would need to perform matching on the pattern fragment of higher-order logic~\cite{Dowek:2001:HUM:778522.778525}. 

Educationally, we plan to perform several experiments concerning the usage of the software in the classroom, as a self-study aid, and as a tool for bridging propositional and first-order reasoning. The application was included in the winter semester 2019 iteration of our introduction to logic course as a laboratory assignment with this particular use case in mind. Analysis of the effect of the application on student understanding and learning is planned for the near future.

As a final remark, while the current problem library is of reasonable size, we plan to continuously update the problem list. We are always looking for new problems to add to the library as well as new categories. The current list reflects necessary problem types for testing the software.

\bibliographystyle{eptcs}
\bibliography{References}

\end{document}